\documentclass[12pt]{article}

\usepackage[margin=1in]{geometry}   
\usepackage{amsfonts}
\usepackage{amsmath} 
\usepackage{booktabs}
\usepackage{multirow}
\usepackage{bbm}
\usepackage{subfig}
\usepackage{relsize}

\usepackage[doublespacing]{setspace} 
\newcommand*{\Scale}[2][4]{\scalebox{#1}{$#2$}}%

\usepackage{float}
\usepackage{caption}
\usepackage{graphicx}
\usepackage{adjustbox}
\usepackage{lscape}

\usepackage{xcolor}

\makeatletter
\newcommand*{\indep}{%
  \mathbin{%
    \mathpalette{\@indep}{}%
  }%
}
\newcommand*{\nindep}{%
  \mathbin{
    \mathpalette{\@indep}{\not}
  }%
}
\newcommand*{\@indep}[2]{%
  \sbox0{$#1\perp\m@th$}
  \sbox2{$#1=$}
  \sbox4{$#1\vcenter{}$}
  \rlap{\copy0}
  \dimen@=\dimexpr\ht2-\ht4-.2pt\relax
  \kern\dimen@
  {#2}%
  \kern\dimen@
  \copy0 
} 
\makeatother

\usepackage{MnSymbol}

\newcommand{\overbar}[1]{\mkern 1.5mu\overline{\mkern-1.5mu#1\mkern-1.5mu}\mkern 1.5mu}

\DeclareMathOperator{\E}{\mbox{E}}

\usepackage{cite}

\definecolor{forestgreen}{RGB}{34,139,34}
\usepackage{hyperref}
\hypersetup{
    colorlinks=true,
    linkcolor=magenta,
    filecolor=magenta, 
    citecolor=green,      
    urlcolor=blue
}

\usepackage{datetime}

\usepackage[utf8]{inputenc}

\makeatletter
\def\@hangfrom#1{\setbox\@tempboxa\hbox{{#1}}%
      \hangindent 0pt
      \noindent\box\@tempboxa}
\makeatother

\usepackage[absolute,showboxes]{textpos}

\TPMargin{5pt}

\newcommand{\copyrightstatement}{
    \begin{textblock}{0.84}(0.08,0.93)    
         \noindent
         \footnotesize
         This draft manuscript presents work in progress. \\
         Comments and reports of mistakes are very much welcome at \href{mailto:issa\_dahabreh@brown.edu}{issa\_dahabreh@brown.edu}.
    \end{textblock}
}

\makeatletter
\def\@seccntformat#1{\@ifundefined{#1@cntformat}%
   {\csname the#1\endcsname\quad}  
   {\csname #1@cntformat\endcsname}
}
\let\oldappendix\appendix 
\renewcommand\appendix{%
    \oldappendix
    \newcommand{\section@cntformat}{\appendixname~\thesection\quad}
}
\makeatother

\usepackage{tkz-graph}
\usepackage{pgf}
\usepackage{tikz}
\usetikzlibrary{arrows,shapes.arrows,shapes.geometric,
  shapes.multipartb,backgrounds,decorations.pathmorphing,positioning,fit,automata}
\tikzset{
  >=stealth',
  true/.style={
    rectangle,
    draw=black, very thick,
    text width=6.;/ mem,
    minimum height=2em,
    text centered,
    fill=gray, opacity = 0.5},
  punkt/.style={
    rectangle,
    rounded corners,
    draw=black, very thick,
    text width=6.5em,
    minimum height=2em,
    text centered},
  est/.style={
    circle,
    draw=black, very thick,
    text centered},
  shade/.style={
    circle,
    draw=black, very thick, fill=gray!50,
    text centered},
  weight/.style={
    circle,
    draw=black, very thick,
    text width=6.5em,
    minimum height=2em,
    text centered},
  pil/.style={
    ->,
    thick,
    shorten <=2pt,
    shorten >=2pt,},
  double/.style={
    <->,
    thick,
    shorten <=2pt,
    shorten >=2pt,},
  dash/.style={
    dashed,
    thick,
    shorten <=2pt,
    shorten >=2pt,},
  dashdouble/.style={
    <->,
    dashed,
    thick,
    shorten <=2pt,
    shorten >=2pt,}
}

\usepackage{authblk}

\def\paperversionmajor{31}
\def\paperversionminor{0}

\begin{document}
\pagenumbering{gobble}

\title{Generalizing causal inferences from randomized trials: counterfactual and graphical identification}

\author[1-4]{Issa J. Dahabreh}
\author[4,5]{James M. Robins}
\author[5]{Sebastien J-P.A. Haneuse}
\author[4-6]{Miguel A. Hern\'an}

\affil[1]{Center for Evidence Synthesis in Health, Brown University School of Public Health, Providence, RI}
\affil[2]{Department of Health Services, Policy \& Practice, School of Public Health, Brown University, Providence, RI}
\affil[3]{Department of Epidemiology, School of Public Health, Brown University, Providence, RI}
\affil[4]{Department of Epidemiology, Harvard T.H. Chan School of Public Health, Harvard University, Boston, MA}
\affil[5]{Department of Biostatistics, Harvard T.H. Chan School of Public Health, Harvard University, Boston, MA}
\affil[6]{Harvard-MIT Division of Health Sciences and Technology, Boston, MA}

\copyrightstatement

\maketitle{}
\thispagestyle{empty}

\clearpage
\thispagestyle{empty}
\vspace*{1in}
\begin{abstract}
\noindent
\linespread{1.3}\selectfont
When engagement with a randomized trial is driven by factors that affect the outcome or when trial engagement directly affects the outcome independent of treatment, the average treatment effect among trial participants is unlikely to generalize to a target population. In this paper, we use counterfactual and graphical causal models to examine under what conditions we can generalize causal inferences from a randomized trial to the target population of trial-eligible individuals. We offer an interpretation of generalizability analyses using the notion of a hypothetical intervention to ``scale-up'' trial engagement to the target population. We consider the interpretation of generalizability analyses when trial engagement does or does not directly affect the outcome, highlight connections with censoring in longitudinal studies, and discuss identification of the distribution of counterfactual outcomes via g-formula computation and inverse probability weighting. Last, we show how the methods can be extended to address time-varying treatments, non-adherence, and censoring.
\end{abstract}

\clearpage
\pagenumbering{arabic}
\section{Introduction}

Randomized trials are widely regarded as the tool of choice to obtain estimates of average treatment effects that are, in expectation, unaffected by baseline confounding. The sample of participants in a trial, however, may not be representative of the \emph{target population} of all trial-eligible individuals, including individuals who meet the trial eligibility criteria but were not invited to participate or those who were invited but chose not to participate. This situation begs the question of whether the effect estimates based on the sample of trial participants can be generalized to the target population.

If individuals in the trial were a random sample of the target population, then the effect estimates from the sample would be directly generalizable to the target population. Unfortunately, that is often not the case. The selection from the target population to the trial sample is a complex process that may result in a different distribution of effect modifiers \cite{dahabreh2016} for individuals in the target population compared with those in the trial sample. When this happens, generalizability requires statistical adjustments and unverifiable assumptions derived from expert knowledge.

In recent years, the problem of selective participation in randomized trials has inspired work  \cite{cole2010, omuircheartaigh2014, tipton2012, tipton2014, zhang2015, buchanan2018generalizing, dahabreh2018generalizing} on ``generalizability methods'' which integrate statistical methods and unverifiable assumptions. However, most developments of generalizability methods (including ours \cite{dahabreh2018generalizing}) have considered simplified settings restricted to time-fixed treatments under complete adherence and have paid little attention to the processes that determine participation in a randomized trial. Specifically, prior work has not distinguished between (1) the invitation to participate in the trial and (2) the decision to participate in the trial after receiving the invitation. A better conceptualization of the selection process is important for two reasons.

First, the mechanisms that determine whether an eligible individual is invited to participate in a trial may differ from those that determine whether an invited individual decides to participate. Therefore, valid generalization from the sample of trial participants to the target population of all trial-eligible individuals will typically require data about these different mechanisms. Though understood in other contexts (e.g., \cite{haneuse2011multiphase}), the need to consider invitation and participation separately has not been considered in the recent literature on generalizability. A practical implication is that generalizability methods will require data from non-invited trial-eligible individuals in addition to data from invited individuals who did and did not participate in the trial.

Second, the invitation to participate in the trial and trial participation itself may have direct causal effects on the outcome that are not mediated by treatment. For example, being made aware of the trial’s existence may affect the behavior of both participants and non-participants (e.g., via Hawthorne effects), and participation may affect components of the participants’ medical care other than the treatment itself. A practical implication is that generalizability methods will need to explicitly consider these direct effects.

In this paper, we structurally describe the invitation and participation components of the selection from the target population of trial-eligible individuals to the sample of randomized trial participants, formalize generalizability from the trial sample to the target population as an attempt to emulate a hypothetical intervention to (1) scale-up trial engagement (both invitation and participation) and (2) assign treatment to the target population. We consider generalizability when trial engagement does and does not directly affect the outcome, highlight connections with censoring in longitudinal studies, and discuss identification of the distribution of counterfactual outcomes via g-formula computation and inverse probability weighting. Lastly, we show how the methods can be extended to address time-varying treatments, non-adherence, and censoring.

\section{Conceptual model}

The \emph{invitation to participate in a trial} is the process that results in certain trial-eligible individuals becoming aware of the trial, being asked to participate in it, and being asked to provide informed consent. \emph{Participation in a trial} is the consent given by some invited individuals to be assigned to a treatment and being followed up as in the trial. We use the term ``engagement'' as shorthand for the combination of the invitation to participate and for participation in the trial.

To fix ideas, consider a randomized trial to compare the effect of two types of surgery, gastric bypass and adjustable gastric band surgery, on weight and quality-of-life after 5 years (similar to the By-Band trial \cite{rogers2014band}). The target population of trial-eligible individuals are adults with a body mass index (BMI) of 40 kg/m$^2$ or more, or BMI of 35 kg/m$^2$ or more and other co-morbidities that could improve with weight loss, and who are fit for anesthesia and surgery in the catchment area of 10 hospitals.

Let $R$ be the indicator of invitation to participate in this trial (1 if invited; 0 otherwise), and $S$ the indicator of trial participation (1 if participating; 0 otherwise). The invitation to participate $R$ is under (partial) investigator control, in the sense that investigators can employ outreach strategies to increase trial awareness among eligible individuals, whereas participation $S$ is almost entirely not under investigator control (except in exceptional circumstances when the requirement for informed consent is waived) \cite{hernan2016discussionkeiding}. 

Let $Z$ be treatment assignment (i.e., 1 for gastric bypass, 0 for adjustable gastric band surgery), and $Y$ the outcome (i.e., weight or quality-of-life at 5 years). Among individuals participating in the trial, treatment assignment occurs by randomization, whereas among non-participants treatment assignment occurs in the form of a recommendation or prescription (the treatments that can be assigned to trial participants and non-participants may be different). Both participants and non-participants in the trial may decide not to adhere to their assigned treatment or the follow-up plan, but, for now, we limit our consideration to trials with perfect adherence complete follow-up, so that the assigned treatment is always the received treatment and there is no drop-out. We revisit non-adherence in Section \ref{sec:time_varying}.

The causal directed acyclic graph (DAG) in panel (A) of Figure \ref{fig:DAG} depicts the causal structure connecting variables $R$, $S$, $Z$, and $Y$. The arrow $R \rightarrow S$ represents that non-invited individuals cannot participate in the trial. The arrow $R \rightarrow  Z$ represents that invited individuals ($R = 1$) who decline participation ($S = 0$) may receive a different treatment recommendation $Z$. The arrow $S \rightarrow Z$ represents that the assignment mechanism is different between trial participants and non-participants. Also, among non-participants, unmeasured prognostic factors $U$ may also affect the treatment assignment, as represented by the fork $Z  \leftarrow  U \rightarrow Y$. Panel (B) of Figure \ref{fig:DAG} shows the causal DAG for the subset of trial participants, in which the following three statements hold: (1) if $S = 1$, then $R = 1$, that is, trial participation implies the invitation to participate; (2) there is no confounding by $U$ of the effect of $Z$ on $Y$; and (3) treatment assignment is not affected by the invitation to participate (no arrow $R \rightarrow Z$). 

Figure \ref{fig:DAG} also includes measured prognostic factors $X$ that may affect the invitation to participate ($X \rightarrow R$), trial participation ($X \rightarrow S$), and treatment assignment ($X \rightarrow Z$), if randomization is conditional on covariates for trial participants ($S=1$) or if treatment recommendations vary depending on the characteristics of non-participants ($S=0$).

A key feature of Figure \ref{fig:DAG} is the consideration of possible direct effects of invitation ($R \rightarrow Y$) and participation ($S \rightarrow Y$) on the outcome that are not through treatment $Z$ \cite{braunholtz2001randomized, peppercorn2004comparison}. That is, we allow for a direct effect of trial engagement on the outcome. 

In our example, invitation $R$ can have a direct effect on the outcome $Y$ if the invitation to participate increases awareness about the adverse effects of obesity, which in turn affects the reporting of body weight even among non-participants \cite{braunholtz2001randomized} or leads to the adoption of healthy behaviors (say, daily exercise) that directly affect body weight (not through $Z$). 

Participation $S$ can have a direct effect on the outcome $Y$ if providers are particularly attentive to trial participants, who in turn are more likely to report improved subjective outcomes (e.g., quality of life) due solely to the increased attention. Effects due to the experience of being observed or experimented on are often referred to as Hawthorne effects \cite{braunholtz2001randomized, landsberger1958}. Another way in which participation $S$ can have a direct effect on the outcome $Y$ is though exposure of the trial participants to concomitant interventions that differ from those in usual practice and that affect the outcome \cite{braunholtz2001randomized}. In our example, participants are required to attend regular clinic visits, during which providers may recommend changes to diet or exercise habits, and these changes may directly affect the outcome.

In the presence of direct effects of trial engagement, generalizability methods need to be modified as described in the next section. As we will see, such effects are hard to identify from studies in which treatment assignment $Z$ is the only randomized intervention \cite{heckman1991randomization}. We will consider settings without direct effects of trial engagement in Section \ref{sec:exclusion_restrictions}.

\section{Joint intervention to scale-up trial engagement and set treatment}

Questions about generalizing causal inferences about treatments assessed in trials can be recast as questions about hypothetical joint interventions to (1) \emph{scale-up trial engagement} to the target population and (2) \emph{assign treatment}. We can use single world intervention graphs (SWIGs) to represent these joint interventions. In Appendix \ref{Appendix_swigs}, we summarize aspects of SWIGs that are relevant to our setup; references \cite{richardson2013primer} and \cite{richardson2013single} provide additional details. Informally, a SWIG can be thought of as a causal DAG depicting a world where we have intervened to set the value of one or more variables in accordance with some treatment regime or strategy. SWIGs \cite{richardson2013single} are attractive graphical tools because they (1) distinguish between nodes that are and are not intervened on and (2) depict counterfactual variables, allowing us to read off independence conditions that involve counterfactuals using d-separation \cite{pearl2009causality, spirtes2000causation}.

Starting with the DAG of Figure \ref{fig:DAG}, we construct the SWIG of Figure \ref{fig:swig_R_S_Z} for a hypothetical intervention to set  $R$ to $r=1$, $S$ to $s=1$ and $Z$ to $z$. We split the nodes $R$, $S$, and $Z$ to denote the joint intervention, inducing three counterfactual outcomes \cite{robins2000d} (potential outcomes \cite{rubin1974}): the participation status under intervention to invite all members of the population to the trial, $S^{r=1}$; the treatment assignment after intervention to scale-up trial participation, $Z^{r=1, s=1}$; and the outcome under joint intervention to scale-up trial engagement and set treatment to $z$, $Y^{r=1, s=1, z}$. Once invitation ($r=1$) and participation ($s=1$) have been intervened on, there are no unmeasured common causes of $Z^{r=1, s=1}$ and $Y^{r=1, s=1, z}$ because under $s=1$ the assignment mechanism is known.

In the next sections we review identifiability conditions and methods to identify the mean outcome under joint interventions on trial engagement and treatment assignment. Of course, identification would be straightforward if all $R$, $S$, and $Z$ could be randomly assigned. Random assignment of $R$ and $S$, however, will be challenging, if not impossible, in most settings. A partial exception was a randomized trial that estimated the effect of obtaining informed consent on outcomes \cite{dahan1986does}: individuals were randomly assigned to provide informed consent and then, regardless of their assignment, were administered a placebo. The trial found that consent had an effect on subjective evaluation of sleep with individuals in the no-consent group reporting ``better hypnotic activity.''

In practice, identification under joint interventions on trial engagement and treatment assignment will necessarily rely on observational data on trial invitation and participation. As a result, the methods described in this paper will yield meaningfully interpretable estimates only if sufficiently well-defined but hypothetical interventions on trial engagement can be proposed and linked to the observed data \cite{hernan2016does}. 

In many cases, those hypothetical interventions may be proposed \cite{holland1986} and, in fact, substantive experts have explicitly discussed the conduct of hypothetical randomized trials where invitation to participate in a trial and trial participation are randomly assigned \cite{peppercorn2004comparison, karjalainen1989treatment}. For example, Peppercorn et al. \cite{peppercorn2004comparison} noted: ``Ideally, the statement that trials are the best treatment option should rest on evidence that trial participants have better outcomes than similar patients treated off-protocol'' and described an ``ethically untenable randomized trial in which patients are randomly assigned (or not) to be offered trial participation.''

In our example, the sharing of information about obesity and strategies for weight reduction communicated to eligible individuals during the invitation to participate in the trial $R$ could be scaled-up (or investigated in a randomized trial). Similarly, any non-protocol-mandated interventions used on trial participants $S$, such as counseling for smoking cessation or blood pressure/diabetes control, can also be scaled-up (or investigated in a randomized trial), alongside the experimental treatment (the surgical approach $Z$).

Also, note that our structural description is by necessity fairly stylized. The invitation to participate $R$ and the participation $S$ in the trial can be subdivided into several components. Each of these components could be represented separately in the causal diagrams. A finer representation of trial engagement, however, does not provide additional conceptual insights so we chose to build our discussion around the compound treatments $R$ and $S$ \cite{hernan2011compound}.

\section{Identification for joint interventions}

\subsection{Identifiability conditions}

Because we are considering the effects of joint interventions on $R$, $S$, and $Z$, we need an expanded version of the ``usual'' identifiability conditions of consistency, exchangeability, and positivity \cite{hernan2019}.

\vspace{0.2in}
\noindent
\textbf{Consistency conditions.} We make the following consistency assumptions for every individual $i$ in the target population,
\begin{align}
&\mbox{if } R_i = 1, \mbox{  then } S^{r=1}_i = S_i; \label{ass:consistency_S} \\
&\mbox{if } R_i = 1 \mbox{ and } S_i = 1,\mbox{  then } Z^{r=1, s=1}_i = Z_i; \label{ass:consistency_Z} \\
&\mbox{if } R_i = 1, S_i = 1, \mbox{ and } Z_i = z, \mbox{ then } Y^{r=1, s =1, z}_i = Y_i. \label{ass:consistency_Y} 
\end{align}
These conditions connect the counterfactual variables with the corresponding observable variables. Condition (\ref{ass:consistency_S}) means that the observed participation status among those actually invited to participate in the trial would be the same as the counterfactual participation under an intervention to scale-up invitation to the target population. Condition (\ref{ass:consistency_Z}) means that the observed treatment assignment among trial participants is the same as the counterfactual assignment under intervention to scale-up trial engagement in the entire population. Condition (\ref{ass:consistency_Y}) means that for trial participants who were actually assigned to treatment $z$, the observed outcome is the same as the counterfactual outcome under the joint intervention on engagement and treatment.

Condition (\ref{ass:consistency_S}) can be reasonable, provided that information about the trial can be provided in a standardized manner. Similarly, condition (\ref{ass:consistency_Z}) is often reasonable, because treatment assignment among trial participants is under the control of the investigators. Condition (\ref{ass:consistency_Y}), however, is somewhat stronger. It entails an assumption of \emph{no hidden versions} of the intervention \cite{rubin1980randomization, rubin2010reflections} or ``treatment variation irrelevance'' \cite{VanderWeele2009}, that is to say, either there is only one way to intervene on $R$, $S$, and $Z$, or the different ways of intervening have the same effect on the outcome. This might not be true when treatments available in the actual randomized trial are impossible to implement without substantial modification when trial engagement is scaled-up. For example, in our motivating example of a trial comparing surgical interventions for obesity, there is some evidence that outcomes under these interventions are better in high-volume centers \cite{courcoulas2003relationship, liu2003characterizing, nguyen2004relationship}. In trials, it is not uncommon to only involve teams from high-volume centers, but outside the confines of the trial, experience is usually more heterogeneous. When scaling-up engagement, such heterogeneity could give rise to outcome-relevant variation in the implementation of the interventions, violating condition (\ref{ass:consistency_Y}). It is possible to extend our results to address multiple versions of treatment along the lines suggested in \cite{hernan2011compound}, provided the versions can be described and assessed (measured) in the data (we return to this issue in the Discussion).

All three consistency conditions entail assumptions of \emph{no interference between units} \cite{ogburn2014causal}; that is to say, that the counterfactual outcomes $S^{r=1}$, $Z^{r=1, s=1}$, and $Y^{r=1, s=1,z}$ of each individual are not affected by the trial engagement and treatment status of any other individual. For condition (\ref{ass:consistency_Z}) this assumption appears fairly mild since it seems plausible that for the $i$th individual actually participating in the trial, the observed treatment assignment is the same as the assignment had we intervened to make them participate. The non-interference assumptions encoded in conditions (\ref{ass:consistency_S}) and (\ref{ass:consistency_Y}), however, can be consequential in certain contexts. For example, consider a randomized trial comparing the effects of a job training program against no training on income. Among those in the trial, individuals assigned to the control arm might obtain and study the materials used in the training group (say, from friends who were assigned to that group); here, the outcome of some non-randomized workers, is influenced by the treatment assignment of randomized workers and condition (\ref{ass:consistency_Y}) fails. In economics, a particular type of interference is termed a ``general equilibrium effect'' \cite{deaton2018understanding}. In the aforementioned job training trial, suppose that the intervention is scaled-up to the entire target population; then, the overall supply of trained workers in the economy would be increased, lessening the impact of the intervention on income.

\vspace{0.2in}
\noindent
\textbf{Exchangeability conditions.} For every $z$,
\begin{align}
Y^{r = 1, s =1, z} &\mathlarger{\mathlarger{\indep}} R \big| X  \label{ass:exchangeability_R}, \\
Y^{r = 1, s =1, z} &\mathlarger{\mathlarger{\indep}} S^{r=1} \big| X, R , \mbox{ and } \label{ass:exchangeability_S} \\
Y^{r = 1, s =1, z} &\mathlarger{\mathlarger{\indep}} Z^{r=1,s=1} \big| X, R, S^{r=1}. \label{ass:exchangeability_Z}
\end{align}

We may view conditions (\ref{ass:exchangeability_R}) through (\ref{ass:exchangeability_Z}) as conditions of \emph{sequential conditional exchangeability}, or perhaps more aptly, \emph{sequential conditional generalizability}. As in the case of time-varying treatments \cite{hernan2019}, the term ``sequential'' highlights that the conditions allow us to generalize, in turn, over invitation, participation, and treatment assignment, conditional on the observed ``past'' at each of these action points. 

Of note, conditions (\ref{ass:exchangeability_S}) and (\ref{ass:exchangeability_Z}), together with conditions (\ref{ass:consistency_S}) and (\ref{ass:consistency_Z}), imply that,
\begin{align}
Y^{r=1, s =1, z} &\mathlarger{\mathlarger{\indep}} S \big| X , R = 1 \mbox{ and } \label{ass:exchangeability_simplified_1}\tag{$5^*$} \\
Y^{r=1, s =1, z} &\mathlarger{\mathlarger{\indep}} Z \big| X , R = 1, S = 1. \label{ass:exchangeability_simplified_2}\tag{$6^*$}
\end{align} 
Condition (\ref{ass:exchangeability_simplified_1}) means that the independence of the counterfactual outcomes $Y^{r=1,s=1,z}$ and trial participation $S$ need only hold among individuals invited to participate in the trial ($R=1$). Condition (\ref{ass:exchangeability_simplified_2}) means that the independence of the counterfactual outcomes $Y^{r=1,s=1,z}$ and treatment $Z$ need only hold among trial participants ($S=1$), \emph{allowing confounding of the $Z \rightarrow Y$ effect in the absence of randomization}. That is to say, when $\displaystyle  Y^{r=1, s =1,z} \mathlarger{\mathlarger{\indep}} Z \big| X , R = 1, S = 1$ holds, $\displaystyle Y^{r = 1 , s =1, a} \mathlarger{\mathlarger{\indep}} Z \big| X , R =1 , S = 0 $ need not hold.

Lastly, readers should bear in mind that the above exchangeability conditions hold for causal structures beyond the one represented by the DAG of Figure \ref{fig:DAG}. For instance, the conditions hold under a new DAG that includes the edges and nodes of Figure \ref{fig:DAG} plus the fork $X \leftarrow U^* \rightarrow Y$ (in this new graph, $U^*$ is a common cause of $X$ and $Y$). This is a general phenomenon that applies to all the causal structures considered in our paper: the counterfactual exchangeability conditions can be viewed as defining an equivalence class of graphs, of which we have chosen to draw only a single member, the simplest one that allows us to focus on the generalizability issues at hand.

\vspace{0.2in}
\noindent
\textbf{Positivity conditions.} We also need the following positivity conditions:
\begin{align}
&\Pr[R =1 | X = x] > 0 \mbox{ for every } x \mbox{ with } f(x) > 0, \label{ass:positivity_R} \\
&\Pr[S =1 | X = x , R = 1] > 0 \mbox{ for every } x \mbox{ with } f(x , R = 1) > 0, \mbox{ and }  \label{ass:positivity_S} \\
&\Pr[Z = z | X = x , R = 1, S = 1] > 0 \mbox{ for every } z \mbox{ and every }  x \mbox{ with }  f(x , R = 1, S = 1) > 0;  \label{ass:positivity_Z} 
\end{align}
where throughout we use $f$ to generically denote densities (the relevant density in each case should be clear from context). Informally, the positivity conditions ensure that at least some individuals who engage in each treatment group of the trial have the covariate patterns needed to ensure the exchangeability conditions in the target population.

\subsection{Identification under joint intervention}

Under the above assumptions, we can identify the counterfactual outcome distribution under joint intervention to set $R$ to $r=1$, $S$ to $s=1$, and $Z$ to $z$:
\begin{equation}\label{identification}
		\Pr[Y^{r=1, s =1, z} \leq y] = \E\! \big[\! \Pr [ Y \leq y | X , R = 1, S =1, Z = z ] \big]. 
\end{equation}
Note here that the expectation on the right-hand-side of the above equation is \emph{with respect to the covariate distribution of the target population}, that is, the population of individuals meeting the trial eligibility criteria (see Appendix \ref{Appendix_joint_intervention_identification} for derivation).

The above identification result pertains to the cumulative distribution function of the counterfactual outcomes, not just the counterfactual outcome mean. In contrast, in prior work we have focused on the identification of just the counterfactual outcome mean \cite{dahabreh2018generalizing}. We are able to obtain a stronger result here because conditions (\ref{ass:exchangeability_R}) through (\ref{ass:exchangeability_Z}) encode assumptions of \emph{exchangeability in distribution}, rather than the assumptions of \emph{exchangeability in mean} invoked in our prior work.

The right-hand side of the above equation, can be re-expressed as
\begin{equation} \label{eq:Gform_IP_equiv}
		\E\! \big[\! \Pr [ Y \leq y | X , R = 1, S =1, Z = z ] \big] = \E \left[ \dfrac{I(Y \leq y , R = 1 , S=1, Z = z) }{ \Pr[R =1, S =1 | X ]  \Pr[Z = z | X, R = 1, S =1 ]  }  \right],
\end{equation}
where $I(\cdot)$ denotes the indicator function. As has been noted before (e.g., see \cite{hernan2019}), we can derive this result without invoking any conditions that involve counterfactuals (we only use the positivity conditions; see Appendix \ref{Appendix_joint_intervention_identification} for an illustration). Thus, the identity holds even when the causal assumptions do not hold, but, in that case, the quantities in (\ref{eq:Gform_IP_equiv}) do not have any causal interpretation. When the assumptions hold, the inverse probability weighting re-expression provides an alternative way for identifying the counterfactual outcome distribution.

\section{When trial engagement does not directly affect the outcome}\label{sec:exclusion_restrictions}

When we can assume that trial engagement effects are negligible, that is to say, the invitation to participate in the trial and trial participation itself do not have a direct effect on the outcome, the DAG of Figure \ref{fig:DAG} is modified by removing the arrows from $R$ and $S$ to $Y$, resulting in the DAG of Figure \ref{fig:DAG_noRS_Y}. 

A randomized trial by Milkman et al.\cite{milkman2011using} provides a good example of a trial in which it is plausible that trial participation does not have any effect on the outcome except through the assigned treatment. Briefly, the investigators wanted to study the impact of different types of mailed prompts on vaccination rates. Using routinely collected insurance information they identified trial-eligible employees in a large firm and randomly assigned them to three treatment groups, which received different prompts about vaccination. Members of the target population were unaware of the eligibility screening process, the requirement for informed consent was waived, there as no indication in the letters that a trial was being conducted, and outcomes were ascertained using routinely collected data sources. As such, in this study it is plausible that there were no direct effects of participation, that is, trial participation affected the outcome only through treatment assignment.

Figure \ref{fig:swig_R_S_Z_noRS_Y} shows the SWIG for joint intervention to set $S$ to $s=1$ and $Z$ to $z$ under the DAG of Figure \ref{fig:DAG_noRS_Y}. Note that this SWIG differs from the one in Figure \ref{fig:swig_R_S_Z} in two ways: first, the arrow from $s=1$ into the counterfactual outcome has been removed, because the arrow was absent in the DAG of Figure \ref{fig:DAG_noRS_Y}. Second, the counterfactual outcome is indexed only by the intervention to set treatment $Z$ to $z$, but it is no longer indexed by the intervention to set $R$ or $S$ (in SWIG construction, this is referred to as \emph{minimal labeling}). In effect, the removal of the $R \rightarrow Y$ and $S \rightarrow Y$ arrows from the DAG reflects \emph{exclusion restriction} assumptions implying that $ Y_i^{r=1, s=1, z} = Y_i^{z}$, for every individual $i$ and every treatment $z$. Under this exclusion restriction, the steps in Appendix \ref{appendix:g_formula_derivation} would not be affected except to substitute  $Y_i^{z} $ for $Y_i^{r=1,s=1, z}; $ thus, we conclude that, under the DAG in Figure \ref{fig:DAG_noRS_Y},
\begin{equation*}
\Pr[Y^{z} \leq y ] = \Pr[Y^{r=1, s=1, z} \leq y] = \E\! \big[\! \Pr [ Y \leq y | X , R = 1 , S =1, Z = z ] \big].
\end{equation*}

\section{Connection with censoring adjustments in time-fixed treatment studies with short follow-up}\label{sec:censoring}

Consider the identification of the counterfactual outcome distribution in randomized or observational studies of time-fixed treatments with longitudinal follow-up. In such studies, individuals who enter the study are often censored (drop-out or are lost-to-follow-up) and identification needs to account for the censoring mechanism. In Appendix \ref{appendix:censoring_point_treatments} we consider a study of a time-fixed treatment with complete adherence to assigned treatment and short follow-up (e.g., a study of patients' food consumption on the 30th post-operative day following bariatric surgery). In such a study, no outcome information will be available for individuals who chose to not complete the study questionnaire and the analysis needs to account for factors that are common causes of loss-to-follow-up and the outcome of interest. 

A natural causal quantity of interest in such a study is the counterfactual outcome distribution under intervention to set $Z$ to $z$ and eliminate censoring \cite{hernan2019}. In Appendix \ref{appendix:censoring_point_treatments}, we show that the conditions needed to identify this quantity and the identification results are strikingly similar to those for identifying the counterfactual outcome distribution under intervention to scale-up trial engagement and set $Z$ to $z$. The similarity should make intuitive sense: generalizability analyses address the issue of individuals \emph{not entering} the trial because they did not know about it or opted not to participate; censoring-adjusted analyses address the issue of individuals \emph{prematurely exiting} the trial.

\section{Time-varying treatments}\label{sec:time_varying}

Up to now, we have discussed time-fixed treatments in studies with complete adherence and short follow-up, such that censoring can be treated as a binary variable. We used this approach because the conceptual issues related to the generalizability of causal inferences from randomized trials to a target population can be illuminated in this simple setting. That said, in realistic applications of generalizability methods the treatments of interest will usually need to be sustained over time, adherence to the assigned treatment will be incomplete, and loss to follow-up will occur at different time points during follow-up (e.g., \cite{westreich2015invited} discuss non-adherence in the context of generalizability analyses). We can readily extend the ideas in the previous sections to address time-varying treatments, including interventions to promote adherence to the assigned treatment or to eliminate censoring in follow-up studies. 

To illustrate this point, we expand our causal structure and introduce some more notation to consider a two-period study, where treatment assignment $Z$ at baseline is followed by an initial treatment period, and where treatment may be sustained (or not) for a second period. Let $A_0$ denote the decision to to receive treatment (or not) during the initial treatment period and $A_1$ the decision to to receive treatment (or not) during the second treatment period. As before, the outcome $Y$ is assessed at the end of the study. Information is collected on time-varying covariates, in addition to the baseline covariates $X$. Specifically, information on time varying covariates $L_0$ is collected after treatment assignment, at the start of the initial treatment period (i.e., between $Z$ and $A_0$) and information on time varying covariates $L_1$ is collected after the initial treatment period, at the start of the second treatment period (i.e., between $A_0$ and $A_1$). Suppose also that the time-varying covariates affect the receipt of future treatment as well as the outcome, and thus are time-varying confounders (see chapter 19 of \cite{hernan2019}). Individuals may be censored at the end of the first treatment period (i.e., before $L_1$ is determined) or the end of the second treatment period (i.e., before $Y$ is determined). Let $C_1$ be the indicator for censoring before the end of the first treatment period and $C_2$ the indicator for censoring before the end of the second treatment period; $C_j = 1$ if censored before the end of the $j$th treatment period, and 0 otherwise, for $j=1,2$. We use overbars to denote histories of random variable that are observed over time; for example, $\overbar{A} = (A_0, A_1)$; $\overbar{L} = (L_0, L_1)$, and $\overbar{C} = (C_1, C_2)$. We use corresponding lowercase symbols to denote realizations of these histories; for example, we denote a specific treatment history as $\overbar{a} = (a_0, a_1)$; a censoring history as $\overbar{c} = (c_1, c_2)$; and the intervention to eliminate censoring as $\overbar{c} = (0, 0) \equiv \overbar{0}$.

Under assumptions analogous to the ones we made earlier in the paper (see Appendix \ref{Appendix_time_varying_tx_identification}), we can identify the distribution of counterfactual outcomes under intervention to scale-up trial engagement, set treatment to $z$, enforce adherence to treatment $\overbar{a} = (a_0, a_1)$, and prevent censoring $\overbar{c} = \overbar{0}$, $\Pr [ Y^{r=1, s=1, z, \overbar{a}, \overbar{c}=\overbar{0}} \leq y ]$, using the g-formula functional or its inverse probability weighting re-expression,
\begin{equation*}
  \begin{split}
        &\Pr [ Y^{r=1, s=1, z, \overbar{a}, \overbar{c}=\overbar{0}} \leq y ] \\ 
        &\enskip = \Scale[0.8]{ \E\! \left[\! \rule{0cm}{0.50cm}  \E \!  \Big[ \E \big[  \Pr [ Y \leq y | X , R = 1, S =1, Z = z , \overbar{L}, \overbar{A} = \overbar{a}, \overbar{C} = \overbar{0} ] \big | X , R = 1, S =1, Z = z, L_0 , A_0 = a_0, C_1 = 0  \big] \Big|  X , R = 1, S =1, Z = z \Big]   \right] } \\
        &\enskip= \E\! \left[\! \dfrac{I(Y \leq y, R = 1, S =1, Z = z, \overbar{A} = \overbar{a}, \overbar{C} = \overbar{0})}{ \Pr[R = 1, S =1 | X] \Pr[Z = z | X, R = 1, S =1] \times  P_{\overbar{a}} \times P_{\overbar{c}} }  \right],
  \end{split}
\end{equation*}
where in the last expression above, we define
\begin{equation*}
  \begin{split}
P_{\overbar{a}} &= \Pr[A_1 = a_1 | X , R = 1, S =1, Z = z, \overbar{L}, A_0 = a_0, C_1 = 0] \\
&\quad \times \Pr[A_0 = a_0 | X , R = 1, S =1, Z = z, L_0 ], \mbox{ and} \\
P_{\overbar{c}} &= \Pr[C_2 = 0 | X , R = 1, S =1, Z = z, \overbar{L}, \overbar{A} = \overbar{a}, C_1 = 0] \\
&\quad \times \Pr[C_1 = 0 | X , R = 1, S =1, Z = z, L_0, A_0 = a_0 ].
  \end{split}
\end{equation*}
The inverse probability weighting re-expression above will be familiar to readers who have studied the identification of the parameters of marginal structural models for time-varying treatments (e.g., see \cite{robins1999association, robins2000a, robins2000b}).

\section{Discussion}

We argue that attempts to generalize inferences from randomized trials to the population of trial-eligible individuals require the notion of interventions to scale-up trial engagement to the population of eligible individuals \cite{dahabreh2019commentaryonweiss}. Prior work on generalizability has nearly exclusively focused on interventions on treatment assignment $Z$ without considering the possibility of intervention on the invitation to participate in the trial $R$ or trial participation $S$. We are only aware of two previous papers that introduced notation for intervention on trial participation (but not on the invitation to participate) \cite{hartman2013, balzer2017all}. Thus, subtleties related to the effects of trial engagement on the outcome have largely remained unappreciated. 

Our approach connects with the broad literature on selective study participation \cite{keiding2016perils} and clarifies the meaning of generalizability analyses when trial engagement directly affects the outcome. Under causal models that allow for such direct effects (e.g., Figure \ref{fig:DAG}), when only intervention on $Z$ is contemplated, it is \emph{not} true that $\displaystyle Y^z \mathlarger{\mathlarger{\indep}} S \big| X$. Yet, this condition (or close variants) has been invoked in prior work \cite{cole2010, kaizar2011, omuircheartaigh2014, tipton2012, tipton2014, zhang2015, buchanan2018generalizing, rudolph2017, dahabreh2018generalizing}. Our results suggest an alternative interpretation for this prior work: in the presence of trial engagement effects, generalizability analyses cannot identify the effect of intervening to just set treatment to a particular level in the target population; but, they can identify the effect of jointly intervening to scale-up the outcome-relevant trial procedures to the entire target population \emph{and} set treatment. When engagement effects are negligible, generalizability analyses can identify the effect of interventions to set treatment, regardless of whether they occur in the context of an experimental study. Whether trial engagement effects exist or not, to generalize causal inferences from a randomized trial to the target population of all trial-eligible individuals, investigators need to obtain covariate data from a representative sample of that population.

Because we wanted to focus on general concepts of generalizability, our exposition was fairly stylized. In particular, we did not spend much time on the policy-relevant work of carefully specifying which components of invitation $R$ and participation $S$ are actually responsible for direct effects on the outcome. Such work will be necessary when scaling-up randomized trials in practice. For example, policy-makers need to consider what information should be provided to individuals eligible for treatment and by what means (accounting for $R$ effects) or whether non-protocol-mandated provider behaviors observed in the trial should also be implemented in the the target population (accounting for $S$ effects).

Bareinboim and Pearl have proposed general methods to assess identifiability and, when possible, to identify the post-intervention distribution in a target population by ``transporting'' information from other sources \cite{pearl2011,bareinboim2012transportability, pearl2014}. The methods rely on DAGs enhanced with selection nodes, but do not address issues related to trial engagement effects because they do not represent interventions on the selection nodes. We chose to focus on the concrete problem of generalizing inferences using trials nested within cohorts of eligible individuals because such studies will be increasingly conducted by embedding pragmatic randomized trials in large health-care systems \cite{staa2014,staa2014a,choudhry2017}. We did not, however, address the somewhat more ambitious goal of extending trial findings to populations ineligible for the trial.

The extension of our results to handle non-adherence and time-varying treatments illustrates that generalizability analyses, which address selective study participation, fit naturally within the ``usual'' causal inference framework in epidemiology and the social sciences: first, select the \emph{causal quantities of interest} that are well-defined in the \emph{target population}. Then, determine the \emph{identifiability conditions} needed in the presence of selective trial participation, confounding, censoring, or non-adherence as assumptions about a sequence of interventions to ``scale-up'' the trial, set treatment, eliminate censoring, and enforce adherence. Next, using these assumptions, find the functionals of the observed data distribution that \emph{identify} the causal quantities of interest. Last, and not addressed in our paper, \emph{estimate the observed data functionals} using appropriate statistical methods \cite{robins1986, robins2000a, bang2005}.

\section{Acknowledgments}
We thank Dr. John Wong (Tufts Medical Center) for pointing us to the randomized trial reported in \cite{dahan1986does}.

This work was supported in part by Patient-Centered Outcomes Research Institute (PCORI) Methods Research Awards ME-1306­-03758 and ME­-1502­-27794, and National Institutes of Health (NIH) grant R37 AI102634. All statements in this paper, including its findings and conclusions, are solely those of the authors and do not necessarily represent the views of the PCORI, its Board of Governors, the Methodology Committee, or the NIH.

\clearpage 
\section{Figures}

\begingroup 
\setlength{\thickmuskip}{0mu}
\begin{figure}[!htbp]
\centering
\caption{DAG depicting the invitation to participate, participation, and treatment assignment in a randomized trial.}\label{fig:DAG}
\vspace{0.2in}
\begin{tikzpicture}[>=stealth, ultra thick, node distance=2cm,
  pre/.style={->,>=stealth,ultra thick,black,line width = 1.5pt}]
  \begin{scope}
  \node [name = X] {$X$};
  \node [name = tagA, above left = 1cm of X] {\textbf{(A)}};
  \node [name = R, right = 2.3cm of X] {$R$};
  \node [name = S, right = 2.3cm of R] {$S$};
  \node [name = Z, right = 2.3cm of S] {$Z$};
  \node [name = Y, right = 2.3cm of Z] {$Y$};
  \node [name = U, below = 1.8cm of Z] {$U$};
  \draw[pil, ->] (0.3,0)  to (2.65,0); 
  \draw[pil, ->] (0.3,0.04)  to[bend right]  (5.55,-0.15); 
  \draw[pil, ->] (0.3,0.04)  to[bend right]  (8.5,-0.15); 
  \draw[pil, ->] (0.3,0.04)  to[bend right]  (11.5,-0.15); 
  \draw[pil, ->] (3.2,0) to (5.55,0); 
  \draw[pil, ->] (3.2,-0.04)  to[bend left]  (8.5,0.15); 
  \draw[pil, ->] (3.2,-0.04)  to[bend left]  (11.5,0.30); 
  \draw[pil, ->] (6.2,0) to (8.5,0); 
  \draw[pil, ->] (9.1,0) to (11.5,0); 
  \draw[pil, ->] (6.2,-0.04)  to[bend left]  (11.5,0.10); 
  \draw[pil, ->] (8.8,-2.1) to (8.8,-0.3); 
  \draw[pil, ->] (8.75,-2.05) to (11.5,-0.35); 
  \end{scope}
  \end{tikzpicture}

  \begin{tikzpicture}[>=stealth, ultra thick, node distance=2cm,
  pre/.style={->,>=stealth,ultra thick,black,line width = 1.5pt}]
  \begin{scope}
  \node [name = X] {$X$};
  \node [name = tagB, above left = 1cm of X] {\textbf{(B)}};
  \node [name = R, right = 2.3cm of X, draw, semithick] {$R = 1$};
  \node [name = S, right = 1.9cm of R, draw, semithick] {$S=1$};
  \node [name = Z, right = 1.85cm of S] {$Z$};
  \node [name = Y, right = 2.3cm of Z] {$Y$};
  \node [name = U, below = 1.8cm of Z] {$U$};
  \draw[pil, ->] (0.3,0)  to (2.65,0); 
  \draw[pil, ->] (0.3,0.04)  to[bend right]  (5.55,-0.15); 
  \draw[pil, ->] (0.3,0.04)  to[bend right]  (8.5,-0.15); 
  \draw[pil, ->] (0.3,0.04)  to[bend right]  (11.5,-0.15); 
  \draw[pil, ->] (3.63,0) to (5.55,0); 
  \draw[pil, ->] (3.63,-0.04)  to[bend left]  (11.5,0.30); 
  \draw[pil, ->] (6.55,0) to (8.5,0); 
  \draw[pil, ->] (9.1,0) to (11.5,0); 
  \draw[pil, ->] (6.55,-0.04)  to[bend left]  (11.5,0.10); 
   \draw[pil, ->] (8.75,-2.05) to (11.5,-0.35); 
  \end{scope}
  \end{tikzpicture}
\end{figure}
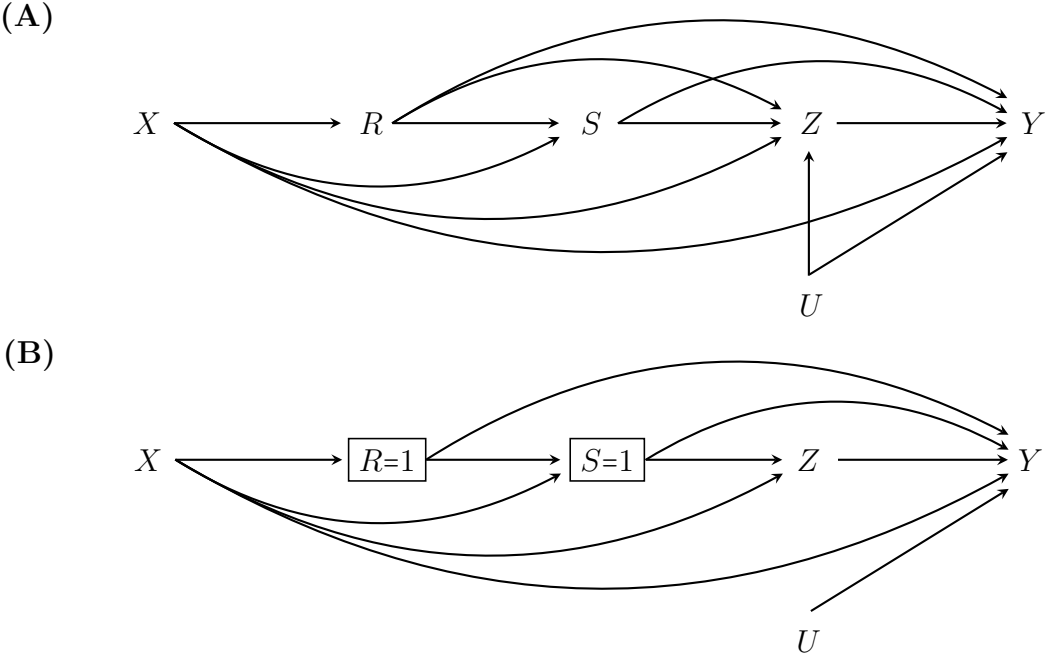
\endgroup

\clearpage
\begingroup 
\setlength{\thickmuskip}{0mu}
\begin{figure}[!htbp]
\caption{SWIG for joint intervention on $R$, $S$, and $Z$.}\label{fig:swig_R_S_Z}
\centering
  \begin{tikzpicture}[>=stealth, ultra thick, node distance=2cm,
  pre/.style={->,>=stealth,ultra thick,black,line width = 1.5pt}]
  \begin{scope}
  \node [name = X] {$X$};
  \node [name = R, right = 1.9cm of X] {$R \enskip \big|  \enskip r = 1$};
  \node [name = S, right = 1.9cm of R] {$S^{r=1} \enskip \big|  \enskip s = 1$};
  \node [name = Z, right = 1.95cm of S] {$Z^{r=1,s=1} \enskip \big| \enskip z $};
  \node [name = Y, right = 2.1cm of Z] {$Y^{r=1,s=1, z}$};
  \draw[pil, ->] (0.3,0)  to (2.3,0); 
  \draw[pil, ->] (0.3,0.04)  to[bend right]  (6.1,-0.2); 
  \draw[pil, ->] (0.3,0.04)  to[bend right]  (10.4,-0.2); 
  \draw[pil, ->] (0.3,0.04)  to[bend right]  (14.8,-0.2); 
  \draw[pil, ->] (4.1,0) to (6.1,0); 
  \draw[pil, ->] (8.4,0) to (10.4,0); 
  \draw[pil, ->] (8.4,-0.04)  to[bend left]  (14.8,0.1); 
  \draw[pil, ->] (4.1,-0.04)  to[bend left]  (10.4,0.2); 
  \draw[pil, ->] (4.1,-0.04)  to[bend left]  (14.8,0.25); 
  \draw[pil, ->] (12.8,0) to (14.8,0); 
  \end{scope}
  \end{tikzpicture}
\end{figure}
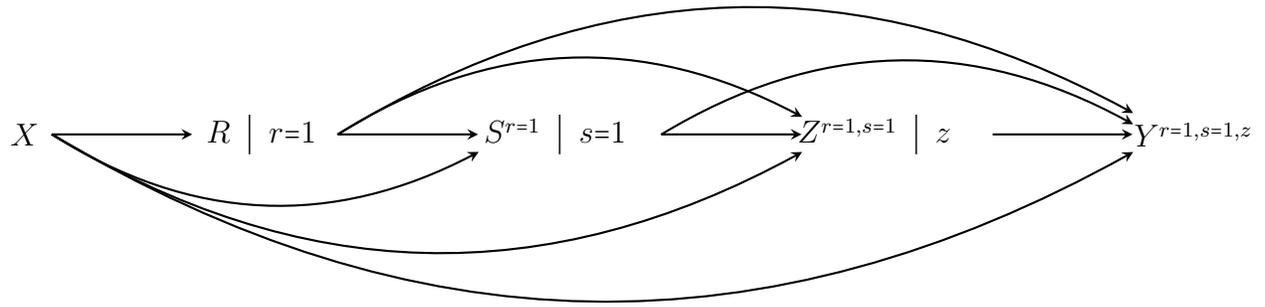
\endgroup

\clearpage

\begingroup 
\setlength{\thickmuskip}{0mu}
\begin{figure}[!htbp]
 \caption{DAG for a trial nested in a cohort study, when trial participation does not affect outcomes.}\label{fig:DAG_noRS_Y}
\centering
  \begin{tikzpicture}[>=stealth, ultra thick, node distance=2cm,
  pre/.style={->,>=stealth,ultra thick,black,line width = 1.5pt}]
  \begin{scope}
  \node [name = X] {$X$};
  \node [name = R, right = 2.3cm of X] {$R$};
  \node [name = S, right = 2.3cm of R] {$S$};
  \node [name = Z, right = 2.3cm of S] {$Z$};
  \node [name = Y, right = 2.3cm of Z] {$Y$};
  \draw[pil, ->] (0.3,0)  to (2.65,0); 
  \draw[pil, ->] (0.3,0.04)  to[bend right]  (5.55,-0.15); 
  \draw[pil, ->] (0.3,0.04)  to[bend right]  (8.5,-0.15); 
  \draw[pil, ->] (0.3,0.04)  to[bend right]  (11.5,-0.15); 
  \draw[pil, ->] (3.2,0) to (5.55,0); 
  \draw[pil, ->] (3.2,-0.04)  to[bend left]  (8.5,0.15); 
  \draw[pil, ->] (6.2,0) to (8.5,0); 
  \draw[pil, ->] (9.1,0) to (11.5,0); 
  \end{scope}
  \end{tikzpicture}
\end{figure}
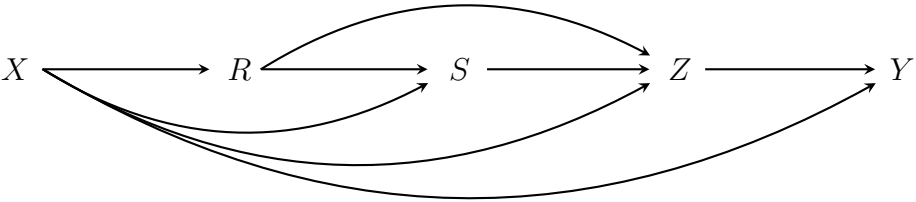
\endgroup

\begingroup 
\setlength{\thickmuskip}{0mu}
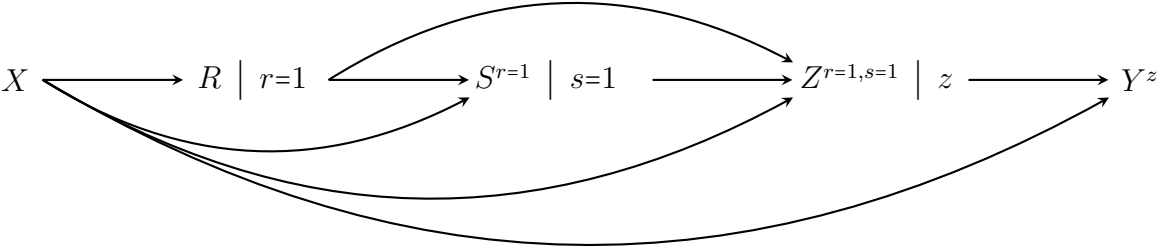
\begin{figure}[!htbp]
 \caption{SWIG for joint intervention on $S$ and $Z$, in the absence of trial engagement effects.}\label{fig:swig_R_S_Z_noRS_Y}
\centering
  \begin{tikzpicture}[>=stealth, ultra thick, node distance=2cm,
  pre/.style={->,>=stealth,ultra thick,black,line width = 1.5pt}]
  \begin{scope}
  \node [name = X] {$X$};
  \node [name = R, right = 1.9cm of X] {$R \enskip \big|  \enskip r = 1$};
  \node [name = S, right = 1.9cm of R] {$S^{r=1} \enskip \big|  \enskip s = 1$};
  \node [name = Z, right = 2.1cm of S] {$Z^{r=1, s=1} \enskip \big| \enskip z $};
  \node [name = Y, right = 1.9cm of Z] {$Y^{z}$};
  \draw[pil, ->] (0.3,0)  to (2.3,0); 
  \draw[pil, ->] (0.3,0.04)  to[bend right]  (6.1,-0.2); 
  \draw[pil, ->] (0.3,0.04)  to[bend right]  (10.4,-0.2); 
  \draw[pil, ->] (0.3,0.04)  to[bend right]  (14.6,-0.2); 
  \draw[pil, ->] (4.1,0) to (6.1,0); 
  \draw[pil, ->] (8.4,0) to (10.4,0); 
  \draw[pil, ->] (4.1,-0.04)  to[bend left]  (10.4,0.2); 
  \draw[pil, ->] (12.6,0) to (14.6,0); 
  \end{scope}
  \end{tikzpicture}
\end{figure}
\endgroup

\clearpage
\bibliographystyle{unsrt}
\bibliography{intervention_on_S}


\ddmmyyyydate 
\newtimeformat{24h60m60s}{\twodigit{\THEHOUR}.\twodigit{\THEMINUTE}.32}
\settimeformat{24h60m60s}
\begin{center}
\vspace{\fill}\ \newline
\textcolor{black}{{\tiny $ $generalizability\_conceptual, $ $ }
{\tiny $ $Date: \today~~ \currenttime $ $ }
{\tiny $ $Revision: \paperversionmajor.\paperversionminor $ $ }}
\end{center}

\clearpage
\appendix


\section{Brief overview of Single World Intervention Graphs (SWIGS)}\label{Appendix_swigs}

\setcounter{table}{0}
\renewcommand{\thetable}{A\arabic{table}}

Starting with a causal DAG about the factual (i.e., observable, even if unmeasured) variables, a SWIG \cite{richardson2013single, richardson2013primer} is obtained by positing interventions on certain nodes; we refer to these nodes as ``intervention nodes.''

First, we ``split'' intervention nodes into a ``random part'' that can be thought as representing the (random) variable in the absence of intervention; and a ``fixed part'' representing the specific intervention under consideration. In this paper, we use a vertical line to denote node splitting. For example, intervention to set $Z$ to $z = 1$ would be depicted as $Z \enskip \big | \enskip z  = 1 $. 

Next, for each split note, incoming arrows on the DAG point into the random part on the SWIG; outgoing arrows on the DAG emanate from the fixed part on the SWIG; otherwise, arrows are left unchanged. 

All descendants of intervention nodes are then relabeled to depict the corresponding counterfactual random variables under the specific intervention under consideration; this is done to signify that, for the descendants of intervention nodes, only post-intervention measurements are possible in the ``world'' where we have intervened in our chosen way. 

Nodes representing random variables can be treated as belonging to a ``conventional'' DAG -- that is to say, the usual rules of d-separation \cite{pearl2009causality,spirtes2000causation} can be used to read-off independence conditions from SWIGs, with the additional property that the fixed nodes block the paths they are intercepting.

\emph{A note on terminology:} Strictly speaking, graphs like the one in Figure \ref{fig:swig_R_S_Z} are ``single world intervention templates'' (SWITs) because by substituting different specific values for $z$ we can use the graph to represent SWIGs under different interventions (formally, a SWIT is a graph-valued function). Because there is no possibility of confusion here, we informally refer to graphs like Figure \ref{fig:swig_R_S_Z} as SWIGs.

\clearpage
\section{Intervention to scale-up trial engagement and set treatment}\label{Appendix_joint_intervention_identification}

\subsection{Identification via g-formula computation}\label{appendix:g_formula_derivation}

\begin{equation*}
  \begin{split}
    \Pr[Y^{r=1, s=1, z} \leq y ] &= \E\! \big[\! \Pr [ Y^{r=1, s=1,z} \leq y | X ] \big] \\
        &= \E\! \big[\! \Pr [ Y^{r=1, s=1,z} \leq y  | X , R =1 ] \big] \\
        &= \E\! \big[\! \Pr [ Y^{r=1, s=1,z} \leq y | X , R = 1, S =1 ] \big] \\
        &= \E\! \big[\! \Pr [ Y^{s=1,z} \leq y | X , R = 1, S =1, Z = z ] \big] \\ 
        &= \E\! \big[\! \Pr [ Y \leq y | X , R = 1, S =1, Z = z ] \big], 
  \end{split}
\end{equation*}
where the first equality follows from the law of total expectation; the second from condition (\ref{ass:exchangeability_R}); the third from condition (\ref{ass:exchangeability_simplified_1}); the fourth from condition (\ref{ass:exchangeability_simplified_2}); and the last from conditions (\ref{ass:consistency_S}) through (\ref{ass:consistency_Y}). Expressions are well defined because of conditions (\ref{ass:positivity_R}) through (\ref{ass:positivity_Z}).

\subsection{Identification via inverse probability weighting}

We can derive the identity in (\ref{eq:Gform_IP_equiv}) as follows:
\begin{equation*}
  \begin{split}
  \E\! \big[ \Pr [Y \leq y | X, R = 1, S = 1, Z = z] \big] &=  \E\! \left[\! \dfrac{\Pr [Y \leq y  | X , S = 1, Z = z] \Pr[R = 1, S = 1, Z = z | X]}{\Pr[R = 1, S = 1, Z = z | X ] }  \right] \\
      &=  \E\! \Bigg[\! \E \left[  \dfrac{ I(Y \leq y, R=1, S  = 1 , Z = z)  }{\Pr[R =1, S =1 | X]  \Pr[Z = z | X, R =1, S =1 ] } \Big| X  \right]  \Bigg]  \\ 
      &= \E\! \left[\!  \dfrac{ I(Y \leq y, R =1, S = 1 , Z = z)}{\Pr[R =1, S =1 | X]  \Pr[Z = z | X, R = 1, S = 1]   }  \right],
  \end{split}
\end{equation*}
where $I(\cdot)$ denotes the indicator function. 

Note that in the above derivation did not use any conditions that involve counterfactual outcomes; thus, the result holds even if $\E\! \big[\! \Pr [Y \leq y | X, R =1, S = 1, Z = z] \big]$ does not have a causal interpretation.

\clearpage
\section{Relationship with identification results in studies with short term follow-up and post-treatment assignment dropout}\label{appendix:censoring_point_treatments}

\setcounter{figure}{0}
\renewcommand{\thefigure}{C\arabic{figure}}

In randomized or observational studies with longitudinal follow-up, individuals who enter the study often drop-out or are lost-to-follow-up. Let $C$ be an indicator for prematurely exiting the study (1 if the individual is censored and 0 if the individual completes the study). A DAG representing the assumption that censoring may be dependent on baseline covariates and treatment is depicted in Figure \ref{fig:DAG_censoring}; note that the DAG also makes the conventional (see Chapter 8 of \cite{hernan2019}) assumption that censoring does not have a direct effect on the outcome (absence of $C \rightarrow$ Y arrow). We have placed a box around the $C$ node to denote that the analysis has to be restricted to individuals who were not censored, because no outcome data is available from censored individuals. The rest of the notation is the same as in previous sections, except that we omit the $S$ node to focus on censoring.

In Figure \ref{fig:swig_Z_C_cens} we show the SWIG generated from Figure \ref{fig:DAG_censoring} for a hypothetical intervention setting treatment $Z$ to $z$ and to eliminate censoring by setting $C$ to $c=0$. We use the exclusion restriction of no $C$-on-$Y$ effect in the DAG, which means that the counterfactual outcomes in the SWIG need only be indexed by $z$ but not $c=0$ (another application of minimal labeling).

\clearpage
\begingroup
\setlength{\thickmuskip}{0mu}
\begin{figure}[!htbp]
\caption{DAG for censoring in a cohort study.}\label{fig:DAG_censoring}
\centering
  \begin{tikzpicture}[>=stealth, ultra thick, node distance=2cm,
  pre/.style={->,>=stealth,ultra thick,black,line width = 1.5pt}]
  \begin{scope}
  \node [name = X] {$X$};
  \node [name = Z, right = 2.3cm of X] {$Z$};
  \node [name = C, right = 2.3cm of Z] {$C$};
  \node [name = Y, right = 2.3cm of C] {$Y$};
  \draw[pil, ->] (0.3,0)  to (2.65,0); 
  \draw[pil, ->] (0.30,0.04)  to[bend right]  (5.55,-0.2); 
  \draw[pil, ->] (0.30,0.04)  to[bend right]  (8.5,-0.15); 
  \draw[pil, ->] (3.2,0) to (5.55,0); 
  \draw[pil, ->] (3.2,-0.04)  to[bend left]  (8.5,0.15); 
  \end{scope}
  \end{tikzpicture}
\end{figure}
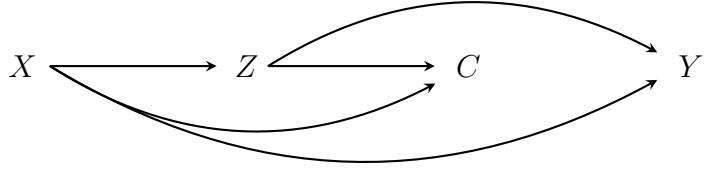
\endgroup

\begingroup 
\setlength{\thickmuskip}{0mu}
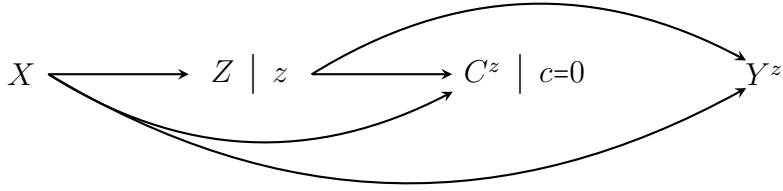
\begin{figure}[!htbp]
\caption{SWIG for joint intervention on $Z$ and $C$.}\label{fig:swig_Z_C_cens}
\centering
  \begin{tikzpicture}[>=stealth, ultra thick, node distance=2cm,
  pre/.style={->,>=stealth,ultra thick,black,line width = 1.5pt}]
  \begin{scope}
  \node [name = X] {$X$};
  \node [name = Z, right = 2cm of X] {$Z \enskip \big|  \enskip z$};
  \node [name = C, right = 2cm of Z] {$C^z \enskip \big|  \enskip c = 0$};
  \node [name = Y, right = 1.8cm of C] {$Y^z$};
  \draw[pil, ->] (0.3,0)  to (2.3,0); 
  \draw[pil, ->] (0.3,0.04)  to[bend right]  (5.8,-0.2); 
  \draw[pil, ->] (0.3,0.04)  to[bend right]  (9.7,-0.15); 
  \draw[pil, ->] (3.8,0) to (5.8,0); 
  \draw[pil, ->] (3.8,-0.04)  to[bend left]  (9.7,0.15); 
  \end{scope}
  \end{tikzpicture}
\end{figure}
\endgroup

We use the following consistency conditions:
\begin{align*}
&\mbox{if } Z_i = z,\mbox{  then } C^{z}_i = C_i; \mbox{ and }  \\  
&\mbox{if } Z_i = z \mbox{ and } C = 0, \mbox{  then } Y^{z,c =0}_i = Y_i. 
\end{align*}
Furthermore, by the exclusion restriction of no $C$-on-$Y$ effect, we have that $Y_i^{z,c=0} = Y_i^{z}$, and we can read-off the following independence conditions from the SWIG of Figure \ref{fig:swig_Z_C_cens}:
\begin{align*} 
Y^{z} \mathlarger{\mathlarger{\indep}} Z \big| X  \mbox{ and } Y^{z} \mathlarger{\mathlarger{\indep}} C^{z} \big| X , Z.
\end{align*} 
Using consistency, the second condition can be written as $\displaystyle Y \mathlarger{\mathlarger{\indep}} C \big| X , Z = z.$

Lastly, we use the following positivity conditions, for every $z$,
\begin{equation*}\label{ass:positivity_cens}
  \begin{split}
&\Pr[Z = z| X = x] > 0, \mbox{ for every } f(x) > 0; \mbox{ and} \\ 
&\Pr[C = 0 | X = x , Z = z] > 0, \mbox{ for every } x \mbox{ such that } f(x , Z = z) > 0.
  \end{split} 
\end{equation*}

\subsection{Identification via g-formula computation}

\begin{equation*}
  \begin{split}
    \Pr[Y^{z, c=0} \leq y ] &= \E\! \big[\! \Pr [ Y^{z, c=0} \leq y | X ] \big] \\
        &= \E\! \big[\!  \Pr [ Y^{z, c=0} \leq y | X , Z =z ] \big] \\
        &= \E\! \big[\! \Pr [ Y^{z, c=0} \leq y  | X , Z = z , C = 0, ] \big] \\ 
        &= \E\! \big[\!  \Pr [ Y \leq y  | X , Z = z , C =0 ] \big], 
  \end{split}
\end{equation*}
where all steps follow from the identifiability conditions above.

\subsection{Identification via inverse probability weighting}

We now show how identification is possible via inverse probability weighting:
\begin{equation*}
  \begin{split}
  \E\! \big[\! \Pr[Y \leq y | X, Z = z, C = 0] \big] &=  \E\! \left[\! \dfrac{\Pr[Y \leq y  | X , Z = z, C = 0]  \Pr[Z = z | X  ]  \Pr[C = 0 | X , Z = z]  }{\Pr[Z = z | X  ]  \Pr[C = 0 | X , Z = z]   }  \right] \\{}
      &=  \E\! \left[\! \dfrac{\E [ I(Y \leq y, Z = z , C = 0)  | X ]}{ \Pr[Z = z | X  ]   \Pr[C = 0 | X , Z = z] }   \right]  \\
      &=  \E\! \Bigg[\! \E \left[  \dfrac{I(Y \leq y, Z = z, C = 0)  }{\Pr[Z = z | X  ]   \Pr[C = 0| X , Z = z]  } \Big| X  \right]  \Bigg]  \\ 
      &= \E \left[\!  \dfrac{I(Y \leq y, Z = z, C = 0)  }{\Pr[Z = z | X  ]   \Pr[C = 0 | X , Z = z ]  }  \right].
  \end{split}
\end{equation*}

Again, the above derivation does not invoke any assumptions that involve counterfactual outcomes; thus, the result holds even when $ \E\! \big[\! \Pr [Y \leq y | X, Z = z, C = 0] \big] $ does not have a causal interpretation.

\clearpage
\section{Time-varying treatments}\label{Appendix_time_varying_tx_identification}

\subsection{Causal quantities of interest and identifiability conditions}

We wish to identify $\Pr[Y^{r=1,s=1,z,\overbar{a}, \overbar{c} = \overbar{0}} \leq y]$, the counterfactual outcome distribution under intervention to scale-up trial engagement ($r=1, s=1$), assign treatment $z$, enforce adherence $\overbar{a} = (a_0, a_1)$, and prevent censoring $\overbar{c} = \overbar{0}$.

Now, suppose that we are willing to make a series of consistency, exchangeability, and positivity conditions that are ``natural'' generalizations of the conditions we used for time-fixed treatments \cite{robins1986, robins1999association, robins2000a,robins2000b}.

\vspace{0.2in}

\noindent
\textbf{Consistency conditions.} For every individual $i$ in the target population, and every $z$, $a_0$, and $a_1$:
\begin{equation*}
\begin{split}
&\mbox{if } R_i = 1, \mbox{  then } S^{r=1}_i = S_i; \\
&\mbox{if } R_i = 1 \mbox{ and } S_i = 1,\mbox{  then } Z^{r=1, s=1}_i = Z_i; \\
&\mbox{if } R_i = 1, S_i = 1, \mbox{ and } Z_i = z, \mbox{ then } L^{r=1, s =1, z}_{0i} = L_{0i}; \\
&\mbox{if } R_i = 1, S_i = 1, \mbox{ and } Z_i = z, \mbox{ then } A^{r=1, s =1, z}_{0i} = A_{0i}; \\
&\mbox{if } R_i = 1, S_i = 1, Z_i = z, \mbox{ and } A_{0i} = a_0, \mbox{ then } C^{r=1, s =1, z, a_0}_{1i} = C_{1i}; \\
&\mbox{if } R_i = 1, S_i = 1, Z_i = z, A_{0i} = a_0, C_{1i} = 0, \mbox{ then } L^{r=1, s =1, z, a_0, c_1 = 0}_{1i} = L_{1i}; \\ 
&\mbox{if } R_i = 1, S_i = 1, Z_i = z, A_{0i} = a_0, C_{1i} = 0, \mbox{ then } A^{r=1, s =1, z, a_0, c_1 = 0}_{1i} = A_{1i}; \\ 
&\mbox{if } R_i = 1, S_i = 1, Z_i = z, \overbar{A} = \overbar{a}, \mbox{ and } C_{1i} = 0, \mbox{ then } C^{r=1, s =1, z, \overbar{a}, c_1 = 0}_{2i} = C_{2i}; \mbox{ and } \\ 
&\mbox{if } R_i = 1, S_i = 1, Z_i = z, \overbar{A} = \overbar{a}, \mbox{ and } \overbar{C} = \overbar{0}, \mbox{ then } Y^{r=1, s =1, z, \overbar{a}, \overbar{c} = \overbar{0}}_{i} = Y_{i}.
\end{split}
\end{equation*}

\clearpage

\noindent
\textbf{Exchangeability conditions.} Under intervention to scale-up trial engagement $r=1$, $s=1$, assign treatment $z$, enforce adherence to treatment $\overbar{a}=(a_0, a_1)$, and prevent censoring $\overbar{c} = \overbar{0}$ (graph not shown), for every $z$, $\overbar{a} = (a_0, a_1)$, $\overbar{c} = (c_1, c_2)$, suppose the following conditions hold:
\begin{equation*}
  \begin{split}
&\quad\Scale[0.9]{Y^{r = 1, s =1, z, \overbar{a},\overbar{c}=\overbar{0}} \mathlarger{\mathlarger{\indep}} R \big| X ; } \\
&\quad\Scale[0.9]{Y^{r = 1, s =1, z, \overbar{a},\overbar{c}=\overbar{0}} \mathlarger{\mathlarger{\indep}} S^{r=1} \big| X, R; } \\
&\quad\Scale[0.9]{Y^{r = 1, s =1, z, \overbar{a},\overbar{c}=\overbar{0}} \mathlarger{\mathlarger{\indep}} Z^{r=1,s=1} \big| X, R, S^{r=1}; }\\
&\quad\Scale[0.9]{Y^{r = 1, s =1, z, \overbar{a},\overbar{c}=\overbar{0}} \mathlarger{\mathlarger{\indep}} A_0^{r=1,s=1,z} \big| X, R, S^{r=1}, Z^{r=1,s=1}, L_0^{r=1,s=1,z}; } \\
&\quad\Scale[0.9]{Y^{r = 1, s =1, z, \overbar{a},\overbar{c}=\overbar{0}} \mathlarger{\mathlarger{\indep}} C_1^{r=1,s=1,z, a_0} \big|  X, R, S^{r=1}, Z^{r=1,s=1}, L_0^{r=1,s=1,z},  A_0^{r=1,s=1,z}; } \\
&\quad\Scale[0.9]{Y^{r = 1, s =1, z, \overbar{a},\overbar{c}=\overbar{0}} \mathlarger{\mathlarger{\indep}} A_1^{r=1,s=1,z, a_0, c_1=0} \big| X, R, S^{r=1}, Z^{r=1,s=1}, L_0^{r=1,s=1,z}, A_0^{r=1,s=1,z}, C_1^{r=1,s=1,z, a_0}, L_1^{r=1,s=1,z, a_0, c_1=0}; \mbox{ and }  }\\
&\quad\Scale[0.9]{Y^{r = 1, s =1, z, \overbar{a},\overbar{c}=\overbar{0}} \mathlarger{\mathlarger{\indep}} C_2^{r=1,s=1,z, \overbar{a}, c_1=0} \big| X, R, S^{r=1}, Z^{r=1,s=1}, L_0^{r=1,s=1,z}, A_0^{r=1,s=1,z}, C_1^{r=1,s=1,z, a_0}, L_1^{r=1,s=1,z, a_0, c_1=0}, A_1^{r=1,s=1,z, a_0, c_1=0}.} 
  \end{split}
\end{equation*}

\clearpage
Note that, when combined with the consistency conditions, the above exchangeability conditions imply 
\begin{equation*}
  \begin{split}
&Y^{r = 1, s =1, z, \overbar{a},\overbar{c} = \overbar{0}} \mathlarger{\mathlarger{\indep}} R \big| X ;  \\
&Y^{r = 1, s =1, z, \overbar{a},\overbar{c} = \overbar{0}} \mathlarger{\mathlarger{\indep}} S \big| X, R = 1;  \\
&Y^{r = 1, s =1, z, \overbar{a},\overbar{c} = \overbar{0}} \mathlarger{\mathlarger{\indep}} Z \big| X, R = 1, S = 1;  \\
&Y^{r = 1, s =1, z, \overbar{a},\overbar{c} = \overbar{0}} \mathlarger{\mathlarger{\indep}} A_0 \big| X, R = 1, S = 1, Z = z, L_0;  \\
&Y^{r = 1, s =1, z, \overbar{a},\overbar{c} = \overbar{0}} \mathlarger{\mathlarger{\indep}} C_1 \big|  X, R = 1, S = 1, Z = z, L_0, A_0 = a_0; \\
&Y^{r = 1, s =1, z, \overbar{a},\overbar{c} = \overbar{0}} \mathlarger{\mathlarger{\indep}} A_1 \big| X, R = 1, S = 1, Z = z, \overbar{L}, A_0 = a_0, C_1 = 0; \mbox{ and }  \\
&Y^{r = 1, s =1, z, \overbar{a},\overbar{c} = \overbar{0}} \mathlarger{\mathlarger{\indep}} C_2 \big| X, R = 1, S = 1, Z = z, \overbar{L}, \overbar{A} = \overbar{a}, C_1 = 0. 
  \end{split}
\end{equation*}

\clearpage
\noindent
\textbf{Positivity conditions.} We also need the following positivity conditions, for every $z$ and every $\overbar{a}$,
\begin{equation*}
  \begin{split}
&\Pr[R =1 | X = x] > 0, \mbox{ for every } x \mbox{ with } f(x) > 0; \\
&\Pr[S =1 | X = x , R = 1] > 0, \mbox{ for every } x \mbox{ with } f(x , R = 1) > 0; \\
&\Pr[Z = z | X = x , R = 1, S = 1] > 0 \mbox{, for every }  x \mbox{ with }  f(x , R = 1, S = 1) > 0; \\
&\Pr[A_0 = a_0 | X = x , R = 1, S = 1, Z = z, L_0 = l_0] > 0, \\
      &\quad\quad\quad\quad\quad \mbox{ for every }  x, l_0 \mbox{ with }  f(x ,l_0 , R = 1, S = 1, Z = z) > 0; \\
&\Pr[C_1 = 0 | X = x , R = 1, S = 1, Z = z, L_0 = l_0, A_0 = a_0] > 0, \\
      &\quad\quad\quad\quad\quad \mbox{ for every }  x, l_0 \mbox{ with }  f(x ,l_0 , R = 1, S = 1, Z = z, A_0 = a_0 ) > 0; \\
&\Pr[A_1 = a_1 | X = x , R = 1, S = 1, Z = z, \overbar{L} = \overbar{l}, A_0 = a_0, C_1 = 0] > 0, \\
      &\quad\quad\quad\quad\quad \mbox{ for every }  x, \overbar{l} \mbox{ with }  f(x , \overbar{l} , R = 1, S = 1, Z = z, A_0 = a_0, C_1 = 0) > 0; \mbox{ and } \\
&\Pr[C_2 = 0 | X = x , R = 1, S = 1, Z = z, \overbar{L} = \overbar{l}, \overbar{A} = \overbar{a} , C_1 = 0] > 0,  \\
      &\quad\quad\quad\quad\quad \mbox{ for every }  x, \overbar{l} \mbox{ with }  f(x , \overbar{l} , R = 1, S = 1, Z = z, \overbar{A} = \overbar{a}, C_1 = 0) > 0.
  \end{split}
\end{equation*}

\clearpage
\subsection{Identification}

\begin{equation*}
  \begin{split}
  &\Scale[0.75]{\Pr[Y^{r=1, s=1, z, \overbar{a}, \overbar{c}=\overbar{0}} \leq y ] } \\
        &\Scale[0.75]{=  \E\! \big[\! \Pr [ Y^{r=1, s=1, z, \overbar{a}, \overbar{c}=\overbar{0}} \leq y | X ] \big] } \\
        &\Scale[0.75]{=  \E\! \big[\! \Pr [ Y^{r=1, s=1, z, \overbar{a}, \overbar{c}=\overbar{0}} \leq y  | X , R =1 ] \big] } \\
        &\Scale[0.75]{=  \E\! \big[\! \Pr [ Y^{r=1, s=1, z, \overbar{a}, \overbar{c}=\overbar{0}} \leq y | X , R = 1, S=1 ] \big] } \\
        &\Scale[0.75]{= \E\! \big[\! \Pr [ Y^{r=1, s=1, z, \overbar{a}, \overbar{c}=\overbar{0}} \leq y | X , R = 1, S=1, Z= z ] \big] } \\ 
        &\Scale[0.75]{= \E\! \Big[\!  \E  \big[   \Pr [ Y^{r=1, s=1, z, \overbar{a}, \overbar{c}=\overbar{0}} \leq y | X , R = 1, S =1, Z = z, L_0  ]   \big |  X , R = 1, S =1, Z = z \big]   \Big] }  \\
        &\Scale[0.75]{= \E\! \Big[\!  \E  \big[ \Pr [ Y^{r=1, s=1, z, \overbar{a}, \overbar{c}=\overbar{0}} \leq y | X , R = 1, S =1, Z = z, L_0 , A_0 = a_0 ]   \big |  X , R = 1, S =1, Z = z \big]   \Big] }  \\
        &\Scale[0.75]{= \E\! \Big[\!  \E  \big[ \Pr [ Y^{r=1, s=1, z, \overbar{a}, \overbar{c}=\overbar{0}} \leq y | X , R = 1, S =1, Z = z, L_0 , A_0 = a_0, C_1 = 0 ] \big |  X , R = 1, S =1, Z = z \big]   \Big]  } \\
        &\Scale[0.75]{= \E\! \left[\! \rule{0cm}{0.50cm} \E  \Big[ \E \big[  \Pr [ Y^{r=1, s=1, z, \overbar{a}, \overbar{c}=\overbar{0}} \leq y | X , R = 1, S =1, Z = z, \overbar{L}, A_0 = a_0, C_1 = 0] \big | X , R = 1, S =1, Z = z, L_0 , A_0 = a_0, C_1 = 0  \big] \Big|  X , R = 1, S =1, Z = z \Big]   \right] }  \\
        &\Scale[0.75]{= \E\! \left[\! \rule{0cm}{0.50cm} \E  \Big[ \E \big[  \Pr [ Y^{r=1, s=1, z, \overbar{a}, \overbar{c}=\overbar{0}} \leq y | X , R = 1, S = 1, Z = z, \overbar{L}, \overbar{A} = \overbar{a}, C_1 = 0] \big | X , R = 1, S =1, Z = z, L_0 , A_0 = a_0, C_1 = 0  \big] \Big|  X , R = 1, S =1, Z = z \Big]   \right] }  \\
        &\Scale[0.75]{= \E\! \left[\! \rule{0cm}{0.50cm}  \E  \Big[ \E \big[  \Pr [ Y^{r=1, s=1, z, \overbar{a}, \overbar{c}=\overbar{0}} \leq y | X , R = 1, S =1, Z = z, \overbar{L}, \overbar{A} = \overbar{a}, \overbar{C} = \overbar{0} ] \big | X , R = 1, S =1, Z = z, L_0 , A_0 = a_0, C_1 = 0  \big] \Big|  X , R = 1, S =1, Z = z \Big]   \right] } \\
        &\Scale[0.75]{= \E\! \left[\! \rule{0cm}{0.50cm}  \E \!  \Big[ \E \big[  \Pr [ Y \leq y | X , R = 1, S =1, Z = z, \overbar{L}, \overbar{A} = \overbar{a}, \overbar{C} = \overbar{0} ] \big | X , R = 1, S =1, Z = z, L_0 , A_0 = a_0, C_1 = 0  \big] \Big|  X , R = 1, S =1, Z = z \Big]   \right] } \\
        &\Scale[0.75]{= \E\! \left[\!   \dfrac{I(Y \leq y, R = 1, S =1, Z = z, \overbar{A} = \overbar{a}, \overbar{C} = \overbar{0})}{\Pr[C_2 = 0, A_1 = a_1 | X , R = 1, S =1, Z = z, \overbar{L}, A_0 = a_0, C_1 = 0] \Pr[C_1 = 0, A_0 = a_0 | X , R = 1, S =1, Z = z, L_0 ] \Pr[R = 1, S =1, Z = z | X]}  \right] } \\
        &\Scale[0.75]{= \E\! \left[\!  \dfrac{I(Y \leq y, R = 1, S =1, Z = z, \overbar{A} = \overbar{a}, \overbar{C} = \overbar{0})}{\Pr[R = 1, S =1 | X] \Pr[Z = z | X, R = 1, S =1] \times P_{\overbar{a}} \times P_{\overbar{c}}  }  \right],}
  \end{split}
\end{equation*}
where in the last expression above, we define
\begin{equation*}
  \begin{split}
P_{\overbar{a}} &= \Pr[A_1 = a_1 | X , R = 1, S =1, Z = z, \overbar{L}, A_0 = a_0, C_1 = 0] \\
&\quad \times \Pr[A_0 = a_0 | X , R = 1, S =1, Z = z, L_0 ], \mbox{ and} \\
P_{\overbar{c}} &= \Pr[C_2 = 0 | X , R = 1, S =1, Z = z, \overbar{L}, \overbar{A} = \overbar{a}, C_1 = 0] \\
&\quad \times \Pr[C_1 = 0 | X , R = 1, S =1, Z = z, L_0, A_0 = a_0 ].
  \end{split}
\end{equation*}

Extensions to multiple time-points, which would be necessary to capture more complex patterns of non-adherence over time, are well-known \cite{robins1986, robins1999association, robins2000a, robins2000b} and do not alter any of our conclusions regarding generalizability to a target population.

\end{document}